# Spurious sources in high-resolution VLA surveys

**Eric F. Jiménez-Andrade** [1,2], **Emmanuel Momjian** [3], **Eric J. Murphy** [2], **James J. Condon** [4] and **Miguel Vega** [1]

[1] *Universidad Nacional Autónoma de México, Instituto de Radioastronomía y Astrofísica., Antigua Carretera a Pátzcuaro 8701, Ex-Hda. San José de la Huerta, 58089, Morelia, Michoacán, México*
[2] *National Radio Astronomy Observatory, 520 Edgemont Road, Charlottesville, VA 22903, USA*
[3] *National Radio Astronomy Observatory, P.O Box O, Socorro, NM 87801, USA*
[4] *Unaffiliated*

**Abstract**

Blind extragalactic radio continuum surveys are increasingly conducted at sub-arcsecond angular resolutions. However, these surveys contain substantially more spurious sources than expected from purely Gaussian noise statistics. We investigate the origin of these false detections using the 10 GHz GOODS-N mosaic obtained with the Karl G. Jansky Very Large Array (VLA), which reaches a native angular resolution of 0.″22 and exhibits a nearly Gaussian pixel-noise distribution. We show that part of the increase in the spurious-source fraction at high angular resolution arises naturally from the larger number of independent resolution elements, while residual wide-field imaging effects associated with the primary-beam response make only a modest contribution. Nevertheless, even after accounting for these effects, the observed fraction of false detections remains several times higher than predicted for Gaussian noise. An analysis of the noise autocorrelation function reveals that images with relatively sparse *uv* coverage retain significant dirty-beam sidelobes that introduce correlated noise fluctuations, substantially increasing the probability of false detections. We conclude that the excess of spurious sources is an intrinsic consequence of incomplete *uv* sampling rather than the imaging methodology itself, highlighting that future deep radio continuum surveys require not only high angular resolution but also sufficiently dense *uv* coverage to minimize spurious detections.

**Resumen**

Los censos extragalácticos de radio continuo están alcanzando resoluciones angulares en el régimen de sub-arcosegundo. Sin embargo, estos censos contienen un número sustancialmente mayor de fuentes espurias del esperado a partir de una estadística de ruido puramente gaussiana. Investigamos el origen de estas detecciones falsas utilizando el mosaico de 10 GHz del campo GOODS-N obtenido con el Karl G. Jansky Very Large Array (VLA), el cual alcanza una resolución angular nativa de 0.″22 y presenta una distribución del ruido por píxel cercana a una distribución gaussiana. Mostramos que parte del incremento en la fracción de fuentes espurias a altas resoluciones angulares se explica de manera natural por el mayor número de elementos de resolución independientes, mientras que los efectos residuales de la generación de imágenes de campo amplio asociados con la respuesta del haz primario representan solo una contribución modesta. No obstante, incluso después de considerar estos efectos, la fracción observada de detecciones falsas permanece varias veces por encima de la predicha para ruido gaussiano. Un análisis de la función de autocorrelación del ruido revela que las imágenes de alta resolución obtenidas con un muestreo relativamente escaso del plano *uv* conservan lóbulos laterales significativos del haz sucio, los cuales introducen fluctuaciones de ruido correlacionadas que incrementan sustancialmente la probabilidad de detecciones falsas. Concluimos que el exceso de fuentes espurias es una consecuencia intrínseca del muestreo incompleto del plano *uv*, más que del algoritmo de síntesis de imagen empleado, lo que indica que los futuros censos profundos de radio continuo requerirán no solo una alta resolución angular, sino también un muestreo del plano *uv* suficientemente denso para minimizar la presencia de detecciones espurias.

**Corresponding author:** Eric F. Jiménez-Andrade *E-mail address:* e.jimenez@irya.unam.mx
**Received:** August 13th, 2026 **Accepted:** October 1, 2026

## 1. INTRODUCTION

Blind extragalactic radio surveys have played a fundamental role in assembling large samples of star-forming galaxies and radio-loud active galactic nuclei detected at GHz frequencies. Historically, these surveys were limited to angular resolutions of $\gtrsim 1''$ (e.g., Becker et al., 1995; Condon et al., 1998; Owen & Morrison, 2008; Owen, 2018; Hale et al., 2021), restricting our ability to probe the detailed radio structures of distant galaxies. However, modern radio interferometers such as the Karl G. Jansky Very Large Array (VLA) have significantly expanded this parameter space. In particular, observations at frequencies above $\sim$ 3 GHz routinely achieve sub-arcsecond angular resolutions while maintaining high sensitivity (e.g., Smolčić et al., 2017; Cotton et al., 2018). Combined with the VLA's wide-field imaging capabilities and improved survey efficiency, these advances have enabled the construction of increasingly deep, high-resolution radio surveys of well-studied extragalactic fields (e.g., Heywood et al., 2021; Jiménez-Andrade et al., 2024).

The scientific impact of these surveys is greatly enhanced when radio catalogs are combined with extensive multiwavelength datasets, providing a powerful framework to investigate the physical processes governing galaxy evolution (e.g., Barro et al., 2019; Weaver et al., 2021). Nevertheless, some of the most intriguing discoveries arise from sources that are detected only at radio wavelengths or in cases where complementary data are unavailable. Establishing the reliability of such detections is therefore crucial. For example, the identification of submillimeter galaxies (e.g., Smolčić et al., 2015; Miettinen et al., 2015, 2017; Thomson et al., 2019) and, more recently, optical and near-infrared (NIR)-dark galaxies (e.g., Talia et al., 2021; van der Vlugt et al., 2023; Gentile et al., 2025) has reshaped our understanding of dust-obscured star formation and galaxy assembly in the early Universe. Since a substantial fraction of cosmic star formation remains hidden by dust out to at least $z \sim 4$ (e.g., Zimmerman et al., 2024; Liu et al., 2025), robust radio source identification is essential for obtaining a complete census of galaxy populations across cosmic time.

The reliability of radio source catalogs can be empirically assessed by estimating the fraction of spurious detections in blind surveys. A common approach consists of identifying positive peaks in the radio image above a chosen signal-to-noise threshold and comparing their abundance with that of negative peaks detected at the same absolute threshold (e.g., Hurley-Walker et al., 2017; Smolčić et al., 2017; van der Vlugt et al., 2021; Boyce et al., 2023). Under the assumption that the image noise is symmetric about zero, the number of negative peaks provides an estimate of the false-positive rate, thereby constraining the fraction of cataloged sources that are likely to be spurious.

An alternative approach is to analytically estimate the expected fraction of spurious detections. Assuming that the image noise has the same angular power spectrum as the synthesized beam, the number of statistically independent noise samples in an image covering an area $\Omega$ can be approximated as $N_{\rm n} \approx 8\ln(2)\Omega/(\pi\theta_{1/2}^2)$, where $\theta_{1/2}$ is the full width at half maximum of the synthesized beam. If the radio image is sensitivity-limited and the distribution of noise amplitudes is well described by a Gaussian function, the probability that any individual noise sample exceeds a threshold of $+5\sigma_{\rm n}$ is $P(> +5\sigma_{\rm n}) = 2.87 \times 10^{-7}$. Consequently, the expected number of false-positive detections above this threshold is given by $N_{\rm n}P(> +5\sigma_{\rm n})$.

This simple analytical framework does not always reproduce the empirically derived fraction of spurious sources reported in blind radio surveys (Murphy et al., 2017). As an increasing number of deep, wide-field radio surveys are conducted at high angular resolution, it becomes essential to quantify the reliability of source catalogs and to characterize the incidence of false detections. In particular, the combination of low detection thresholds and small synthesized beams may substantially affect the fraction of spurious sources, with important implications for the identification and interpretation of faint radio populations.

Here, we exploit the 10 GHz mosaic of the Great Observatories Origins Deep - North (GOODS-N) field (Jiménez-Andrade et al., 2024), which reaches a depth of $\sigma_{\rm n} = 671$ nJy beam$^{-1}$ at an angular resolution of 0.″22, as a unique testbed to investigate how the synthesized beam size influences the fraction of spurious sources and to assess whether empirical measurements agree with analytical predictions. In Section 2, we describe the 10 GHz GOODS-N mosaic and the methodology adopted in this work. In Section 3, we quantify the dependence of the spurious source fraction on angular resolution. Finally, in Section 4, we discuss the implications of our findings and summarize our conclusions.

## 2. DATA

The 10 GHz GOODS-N survey is based on 380 hr of VLA observations obtained in the A, B, and C configurations (Project code: VLA 16B-320; Principal Investigator: Eric J. Murphy) using the X-band (8–12 GHz) receivers. 17 individual pointings were arranged to provide nearly uniform sensitivity across the central region of the field. The observations cover a total area of 297 arcmin$^2$ centered at J2000 R.A. = $12^{\rm h}36^{\rm m}55^{\rm s}$ and Decl. = $+62°14'15''$. The final full-resolution mosaic reaches a median rms sensitivity of $\sigma_{\rm n} = 671$ nJy beam$^{-1}$ with a synthesized beam of 0.″22 and pixel scale of 0.″05/pixel, making it the deepest and highest-angular-resolution extragalactic radio survey conducted at frequencies above 3 GHz. An additional mosaic was produced to maximize the brightness sensitivity, yielding a synthesized beam of $\theta_{1/2} = 1.''0$, pixel scale of 0.″25/pixel, and an rms sensitivity of $\sigma_{\rm n} = 968$ nJy beam$^{-1}$. Details of the data reduction, calibration procedures, imaging strategy, and source catalog construction are presented in Jiménez-Andrade et al. (2024)[1].

In Jiménez-Andrade et al. (2024), we noted that the fraction of negative detections, and consequently the inferred fraction of spurious sources, appeared to be substantially lower in maps constructed at a coarser angular resolution. This finding motivated the present work, in which we systematically quantify how the fraction of spurious sources varies with the angular resolution of the GOODS-N 10 GHz mosaic. Given the large number of synthesized beams encompassed by the survey area and the low detection thresholds required to probe the faint radio population, this dataset provides an ideal testbed for assessing the reliability of source catalogs derived from deep, high-resolution radio observations.

## 3. DATA ANALYSIS AND RESULTS

### 3.1. Empirical Determination of the Spurious Source Fraction

To investigate how the fraction of spurious sources depends on angular resolution, we generated a series of lower-resolution mosaics from the non-primary-beam-corrected GOODS-N 10 GHz mosaic, which has a synthesized beam of $\theta_{1/2} = 0.''22$. Using the CASA task `imsmooth`, the image was convolved to angular resolutions of 0.″25, 0.″3, 0.″4, 0.″5, 0.″6, 0.″8, and 1.″0. A uniform pixel scale of 0.″05/pixel was adopted for all the images. Starting from the non-primary-beam-corrected mosaic preserves the nearly uniform noise distribution across the field, producing "flat-noise" images at all resolutions. This avoids the highly elevated noise levels present near the edges of primary-beam-corrected mosaics and enables a more robust determination of the local rms noise during source extraction. Sources were identified using `PyBDSF` with an island detection threshold of `thresh_isl` $= 3\sigma$, where $\sigma$ is the local rms noise, and a peak detection threshold of `thresh_pix` $= 5\sigma$.

To estimate the number of spurious detections, source extraction was repeated on inverted versions of each mosaic obtained by multiplying the images by $-1$. The same `PyBDSF` detection parameters were used for both the original and inverted images. As demonstrated in Figure 1, the pixel brightness distributions of the mosaics are well described by zero-centered

[1] https://science.nrao.edu/science/surveys/vla-x-gn/home

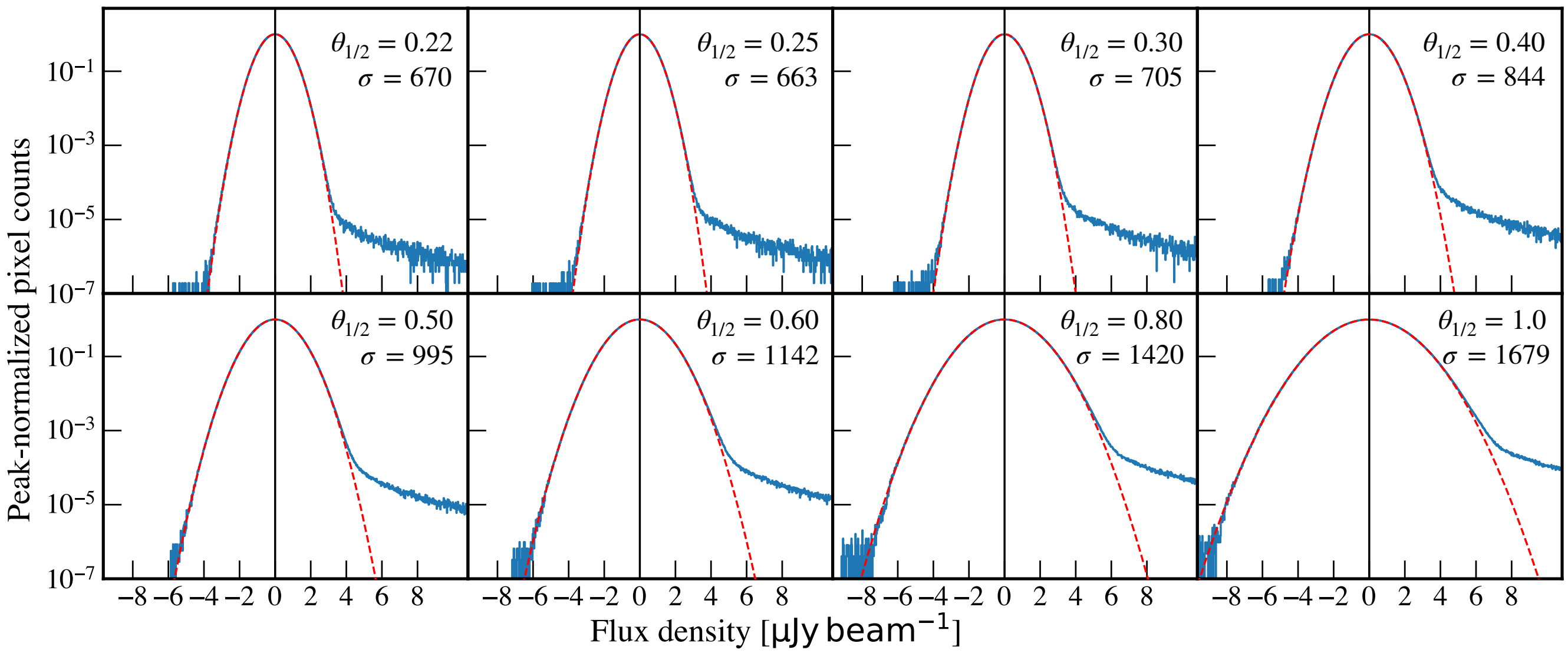

**Figure 1.** Pixel brightness distributions of the GOODS-N 10 GHz mosaics (before primary-beam correction) at different angular resolutions. For visualization purposes, each histogram was divided by its maximum value, such that the peak of each distribution is unity. The upper-right corner of each panel lists the angular resolution, $\theta_{1/2}$ (arcsec), and the image rms noise, $\sigma$ (nJy beam$^{-1}$). The positive excess corresponds to real radio sources, while no significant negative excess is observed. The red dashed line shows a Gaussian fit, demonstrating that the image noise is well described by a symmetric Gaussian distribution centered at zero. Under these conditions, negative emission peaks provide a robust empirical estimate of the spurious source population arising from random noise fluctuations.

Gaussian functions, indicating that the image noise is statistically symmetric about zero. Under this assumption, negative detections provide an empirical estimate of the number of false-positive sources expected in the corresponding positive image. The spurious source fraction is therefore measured as a function of angular resolution as $f_{\rm ss} = N_{\rm neg}/N_{\rm pos}$, where $N_{\rm pos}$ and $N_{\rm neg}$ are the numbers of positive and negative detections, respectively, with peak brightness greater than the detection threshold of $5\sigma_{\rm n}$.

Finally, we verify that the values of $N_{\rm pos}$ and $N_{\rm neg}$, and therefore $f_{\rm ss}$, are not significantly affected by the method used to generate the lower-resolution mosaics. Specifically, consistent results are obtained when the images are produced through $uv$-tapering rather than by convolving the final mosaic with `imsmooth`. The primary advantage of the latter approach is computational efficiency, as $uv$-tapering requires re-imaging and deconvolution of the visibility data for each angular resolution. We further verify that the measured values of $f_{\rm ss}$ are largely insensitive to the source extraction algorithm by repeating the analysis with `Blobcat`, obtaining results consistent with those derived using `PyBDSF`.

### 3.2. Spurious Source Fraction as a Function of Angular Resolution

Figure 2 shows the fraction of spurious sources as a function of peak signal-to-noise ratio (SNR) for the eight GOODS-N 10 GHz mosaics with angular resolutions ranging from 0.″22 to 1.″0. As expected, the vast majority of spurious detections occur at SNR $\lesssim$ 6, consistent with the pixel brightness distributions shown in Figure 1. Those distributions are well described by zero-centered Gaussian functions and exhibit no significant excess in their negative tails, indicating that the image noise is close to Gaussian and that the mosaics are largely free of bright positive imaging artifacts producing high-significance false detections.

The nearly Gaussian nature of the noise allows us to compare the observed number of spurious detections with the analytical expectation for Gaussian random fields. For Gaussian noise, the expected number of false detections above a threshold of $+5\sigma_{\rm n}$ is

$$N_{\rm neg} = N_{\rm n} P(> +5\sigma_{\rm n}) = N_{\rm n}\,(2.87 \times 10^{-7})\,, \quad (1)$$

where

$$N_{\rm n} \approx \frac{8\ln(2)\,\Omega}{\pi\theta_{1/2}^2} \quad (2)$$

is the number of independent noise elements within an image of solid angle $\Omega$. Note that, because this calculation is based on the number of independent beam-sized resolution elements, Equation 1 describes the expected number of beam-scale noise peaks, and therefore does not account for the effects of extended source morphology on the probability of a spurious source detection.

Since the fraction of spurious sources is

$$f_{\rm ss} = \frac{N_{\rm neg}}{N_{\rm real}} \propto \frac{N_{\rm n}}{N_{\rm real}} \propto \frac{1}{\theta_{1/2}^2 N_{\rm real}}, \quad (3)$$

its value is expected to decrease toward coarser angular resolutions, simply because the number of statistically independent noise elements decreases as the synthesized beam becomes larger. Consequently, a direct comparison of $f_{\rm ss}$ between images with different angular resolutions is affected by this purely statistical scaling.

To isolate any intrinsic dependence on angular resolution, in the following we use the spurious-source fraction normalized by the number of independent noise elements, $f_{\rm ss}/N_{\rm n}$. This quantity removes the trivial dependence on the number of independent resolution elements, allowing a more meaningful comparison between images with different synthesized beams.

Figure 3 compares the measured values of $f_{\rm ss}/N_{\rm n}$ with the expectation for purely Gaussian noise. If Gaussian statistics alone determined the incidence of false detections, the normalized spurious fraction would closely follow the theoretical prediction (upper panel) and the ratio $f_{\rm ss}^{\rm obs}/f_{\rm ss}^{\rm exp}$ would remain close to unity (lower panel). Instead, we find that the observed values systematically exceed the Gaussian expectation at all angular resolutions. At the highest resolution ($\theta_{1/2} = 0\farcs22$), the observed spurious fraction per independent noise element is approximately nine times larger than predicted, while this excess gradually decreases with increasing beam size, reaching a factor of $\sim 4$ at $\theta_{1/2} = 1\farcs0$. This suggests that the excess of spurious sources in high-resolution radio images cannot be explained solely by Gaussian noise fluctuations, but instead reflects the presence of additional interferometric imaging effects.

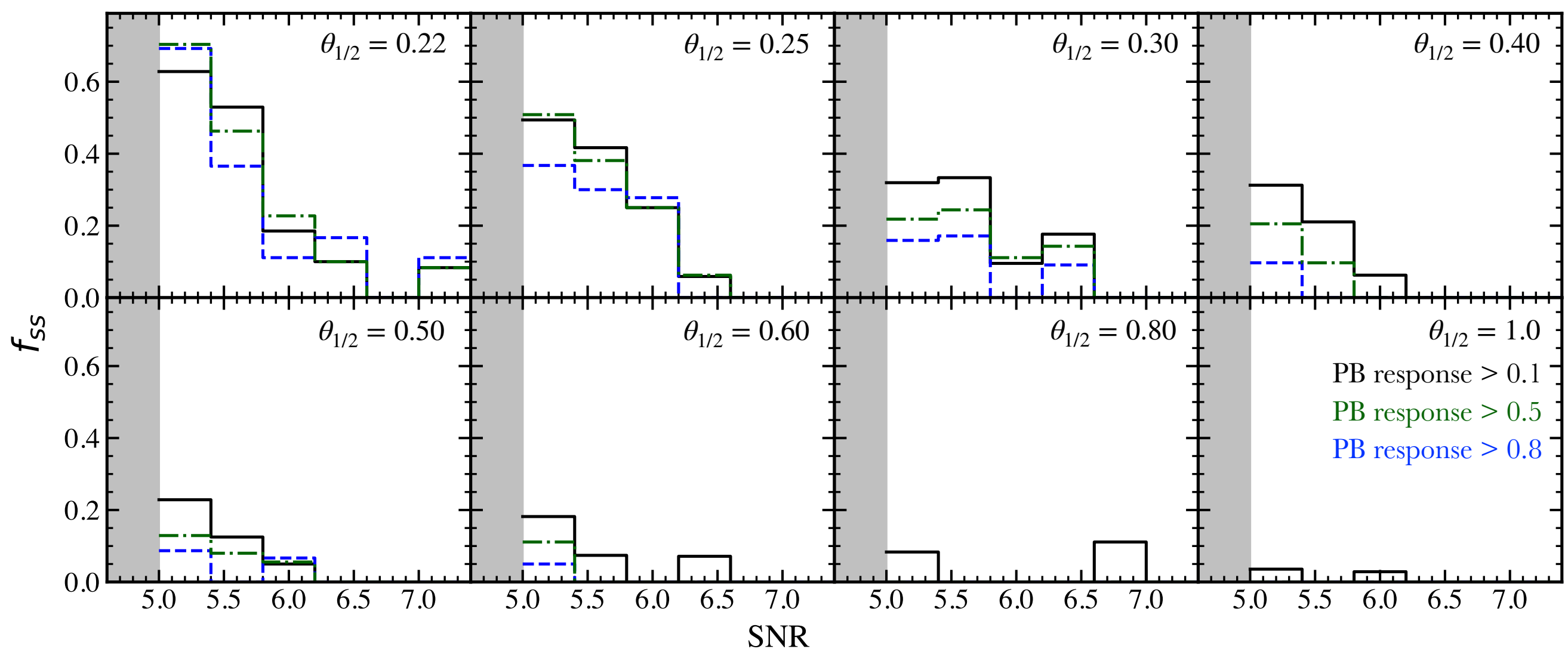

**Figure 2.** Fraction of spurious detections, $f_{\rm ss}$, as a function of peak signal-to-noise ratio (SNR) in the GOODS-N 10 GHz mosaics at different angular resolutions. Here, $f_{\rm ss}$ is defined as the number of spurious detections divided by the total number of detections within each SNR bin and is therefore a differential, rather than cumulative, quantity. The blue dashed and green dash-dotted histograms show the spurious fractions derived from regions where the primary-beam (PB) response exceeds 80% and 50%, respectively, while the black histograms show the fractions measured over the full mosaic, corresponding to the area where the primary-beam response exceeds 10%. As expected from the nearly Gaussian and symmetric noise distributions shown in Figure 1, spurious detections are concentrated at low SNR ($\lesssim 6$). A clear decline in $f_{\rm ss}$ is observed as the image is smoothed to progressively coarser angular resolutions, demonstrating that image smoothing significantly reduces the contamination of source catalogs by noise-induced false positives.

### 3.3. Effects of the Imaging Methodology and Primary-Beam Response

To determine whether the elevated spurious fraction depends on the adopted imaging strategy, we re-imaged one of the 17 individual VLA pointings that comprise the GOODS-N mosaic. Restricting the analysis to a single pointing reduces the computational cost while preserving the relevant imaging characteristics, allowing us to explore several imaging strategies commonly used to improve wide-band and wide-field interferometric imaging.

Specifically, we produced a set of $0\farcs22$ images using `tclean` in CASA, adopting four different imaging configurations. All configurations employed `gridder='widefield'`, `specmode='mfs'`, and `weighting='briggs'` with `robust=2.0`. The four imaging strategies were: (1) `deconvolver='mtmfs'` (multi-term, multi-frequency synthesis (MT-MFS); Rau & Cornwell, 2011) combined with $w$-projection (Cornwell et al., 2005), corresponding to the imaging approach adopted for the original GOODS-N mosaics; (2) $w$-projection with the default `deconvolver='hogbom'`; (3) facetted imaging (with `facets=3`) with the default `deconvolver='hogbom'`; (4) `deconvolver='mtmfs'` without $w$-projection; and (5) the default `deconvolver='hogbom'` without $w$-projection. The MT-MFS algorithm models the frequency dependence of the sky brightness across the observed bandwidth, thereby reducing spectral deconvolution artifacts and improving image fidelity in wide-band observations. In contrast, $w$-projection and facetted imaging corrects/account for non-coplanar baseline ($w$-term) effects that become important in wide-field imaging, mitigating image distortions away from the phase center.

For each imaging configuration, we measured the fraction of spurious sources using the same source-extraction procedure described in Section 3.1. The measured spurious fractions range from 0.25 to 0.40, although all values are mutually consistent within their $1\sigma$ uncertainties (Figure 4). No statistically significant dependence on the adopted imaging strategy is observed, indicating that neither the treatment of spectral variations nor the correction of non-coplanar baseline effects has a significant impact on the inferred fraction of false detections.

Since wide-field imaging errors are generally expected to become more severe away from the phase center, we investigate whether the spatial distribution of spurious detections depends on the primary-beam (PB) response. We therefore compare $f_{\rm ss}/N_{\rm n}$ measured from the GOODS-N 10 GHz mosaics at different angular resolutions using only the regions enclosed by the 80% and 50% PB response levels (Figure 3). Restricting the analysis to progressively higher PB response regions results in a reduction in the normalized spurious-source fraction at all angular resolutions. Since the mosaics are not corrected for the primary-beam response, the image noise remains approximately uniform across the field, implying that this reduction is unlikely to arise solely from variations in the thermal noise. Instead, residual wide-field imaging artifacts—including imperfect corrections for non-coplanar baselines and other direction-dependent effects—are expected to become increasingly significant toward the edges of the mosaic, even when modern imaging techniques such as $w$- projection are employed. In addition, pointing errors introduce larger fractional amplitude errors toward the edges of the primary beam.

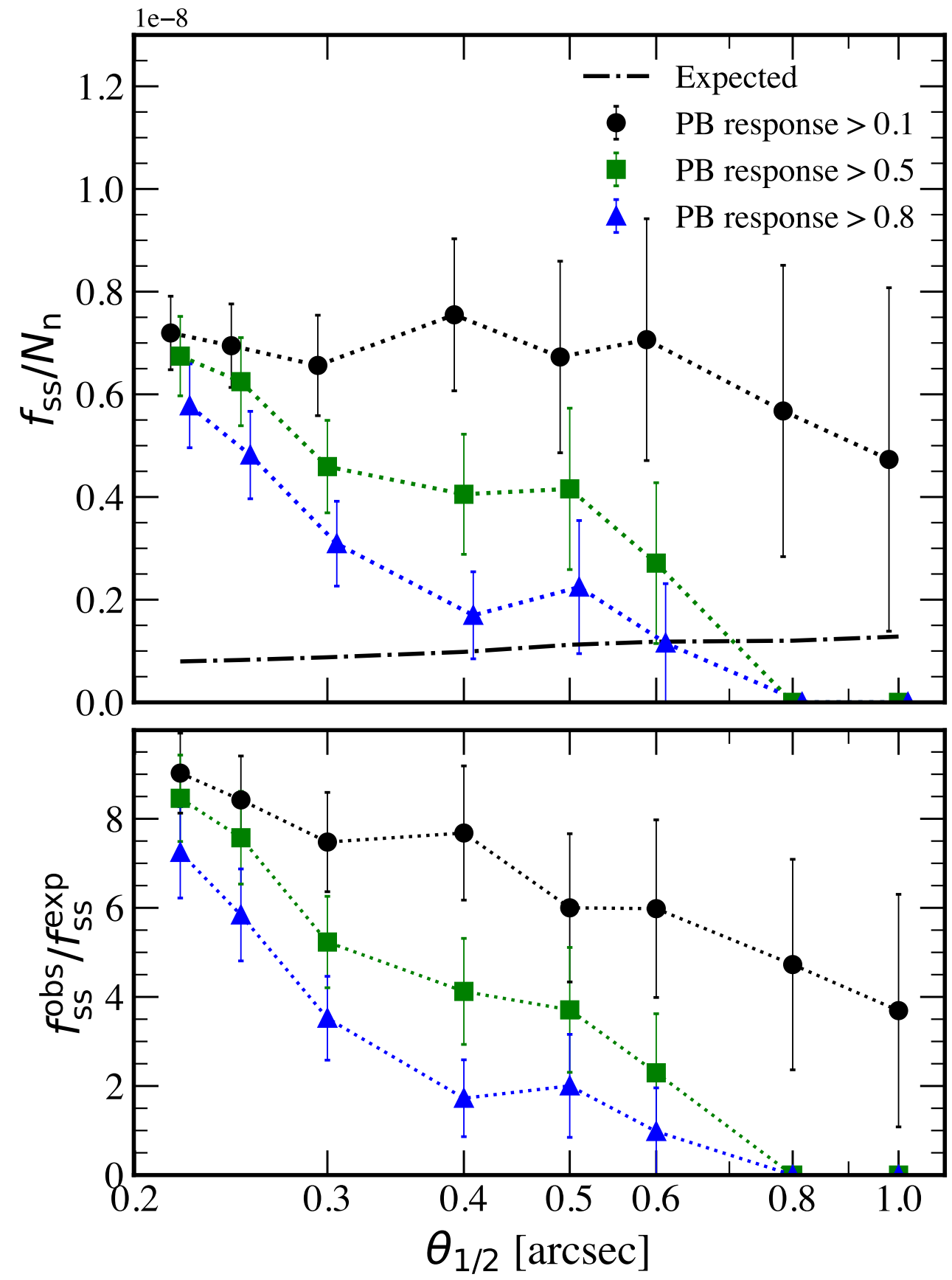

**Figure 3.** *Top panel:* Measured normalized spurious-source fraction, $f_{ss}/N_n$, as a function of synthesized beam FWHM for three primary-beam (PB) response thresholds (PB > 0.1, 0.5, and 0.8). The dash-dotted line shows the Gaussian expectation. Restricting the analysis to regions with higher PB response reduces the normalized spurious-source fraction, but a significant excess above the Gaussian prediction persists at the highest angular resolutions. *Bottom panel:* Ratio of the observed to the expected values of $f_{ss}/N_n$. Although the discrepancy decreases when only high-PB regions are considered, the observed values remain up to a factor of $\sim 7$ larger than expected at the native angular resolution, demonstrating that residual wide-field imaging effects alone cannot account for the excess of spurious detections.

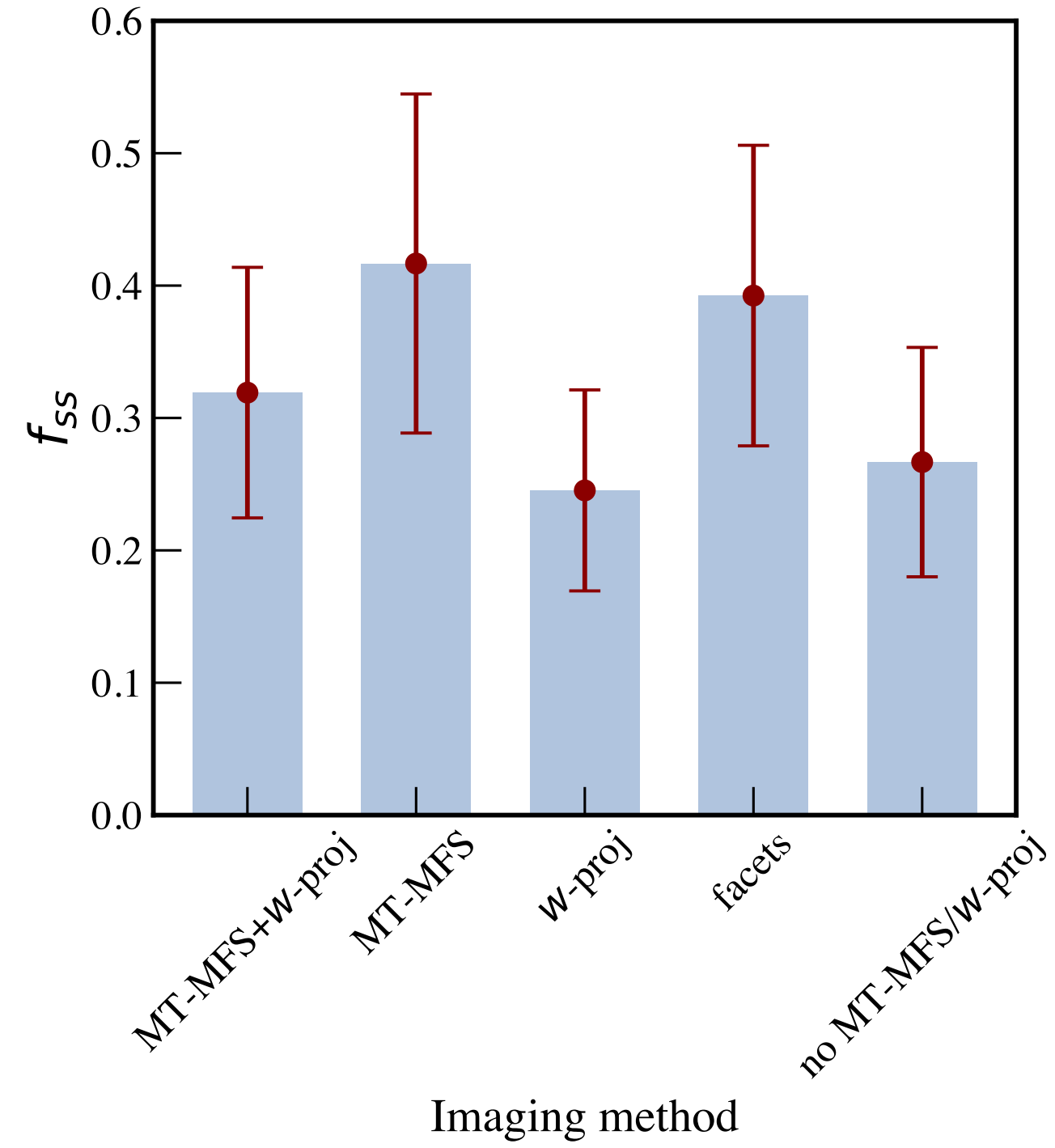

**Figure 4.** Total fraction of spurious sources, $f_{ss}$, measured for radio images with an angular resolution of 0.″22 produced using different imaging strategies. For this test, data from a single pointing of the 17-pointing GOODS-N 10 GHz mosaic were used. From left to right, the imaging approaches are: MT-MFS plus *w*-projection, MT-MFS only, *w*-projection only, facetted imaging only, and neither MT-MFS nor *w*-projection. Error bars correspond to Poisson uncertainties. The measured values of $f_{ss}$ are statistically consistent across all imaging methods, indicating that the prevalence of spurious detections is not strongly affected by the choice of deconvolution or wide-field imaging algorithm.

Although excluding the outer regions of the mosaic lowers the overall fraction of spurious detections, the improvement is relatively modest. For example, at the native angular resolution of 0.″22, restricting the analysis to the region with PB response greater than 80% reduces the normalized spurious-source fraction, $f\text{ss}/N_n$, by approximately 20% relative to the full mosaic (top panel of Figure 3). Likewise, the discrepancy between the observed and expected spurious-source fractions is reduced, but remains substantial (bottom panel of Figure 3): at 0.″22, $f_{ss}$ is still approximately seven times larger than expected, even when only the highest-PB region is considered. Furthermore, the decline of $f_{ss}$ with increasing angular resolution is nearly identical for all three PB cuts, indicating that the strong dependence on angular resolution is largely independent of position within the mosaic. Thus, while residual wide-field imaging effects contribute to the overall population of spurious detections, they cannot explain either the strong dependence of $f_{ss}$ on angular resolution or the substantial discrepancy between the expected and observed values of $f_{ss}$ at the highest angular resolutions. These findings instead point to a more fundamental property of interferometric imaging, which we discuss in the following section.

## 4. DISCUSSION AND CONCLUSIONS

A fundamental limitation of radio interferometry is the incomplete sampling of the *uv* plane. Sparse or non-uniform *uv* coverage produces correlated noise and residual sidelobe structures. Consequently, the fraction of spurious detections might depend primarily on the statistical properties of the interferometric point-spread function (PSF) and the underlying *uv* sampling, rather than on the imaging software used to reconstruct the image.

To investigate this interpretation, we measured the noise autocorrelation function (ACF) of two versions of the GOODS-N 10 GHz mosaic, with angular resolutions of 0.″22 and 1.″0. The ACF, tracing the dirty beam (i.e., the PSF), provides a direct characterization of the spatial correlation of the image noise and has recently been shown to be a powerful diagnostic of correlated noise in interferometric images (Tsukui et al., 2023). We computed the ACF using the Python package `ESSENCE` (Evaluating Statistical Significance under Noise Correlation;

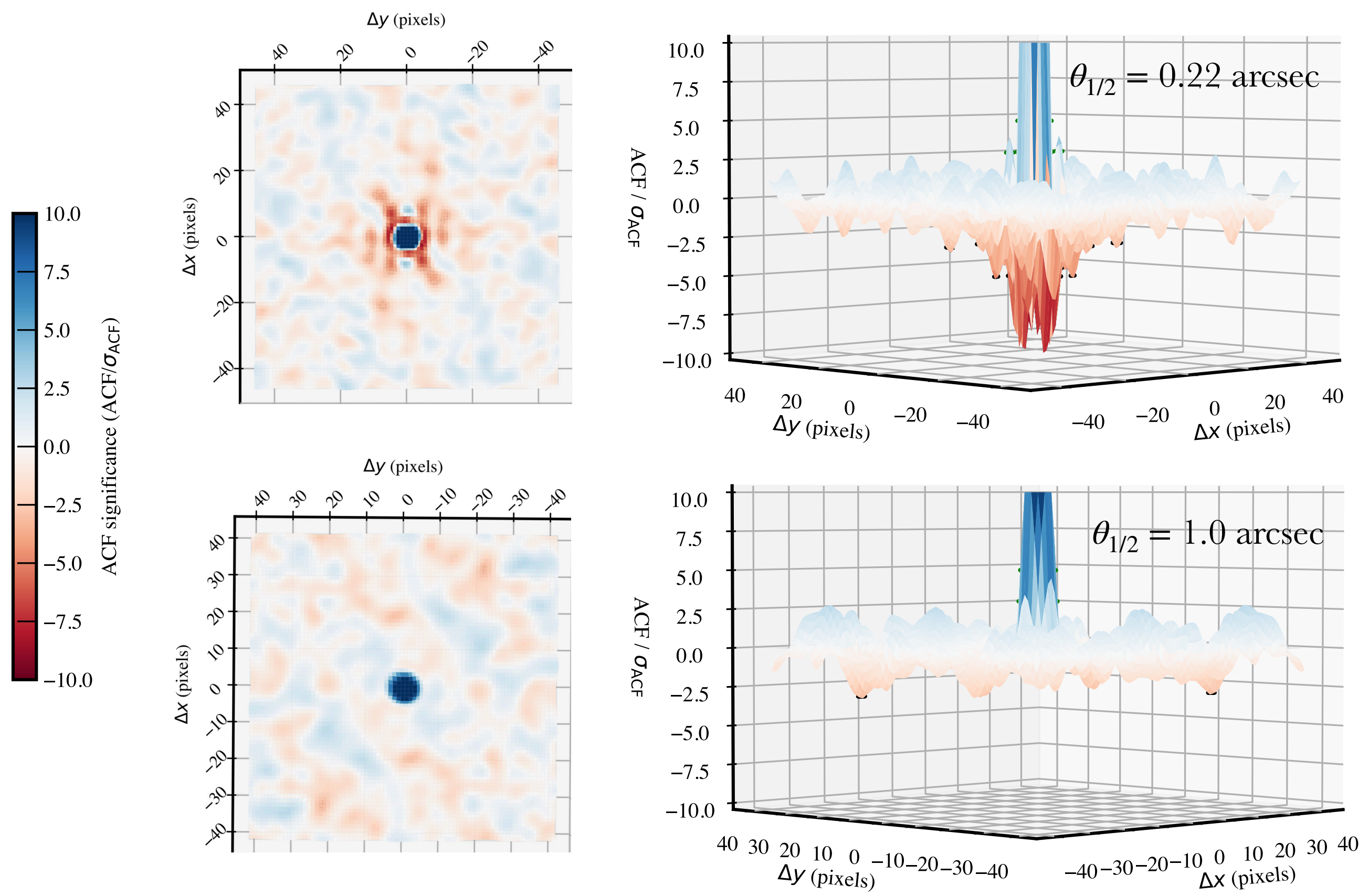

**Figure 5.** Significance of the autocorrelation function ($ACF/\sigma_{ACF}$) of the GOODS-N 10 GHz mosaic at angular resolutions of $\theta_{1/2} = 0\farcs22$ (top) and $\theta_{1/2} = 1\farcs0$ (bottom). The left panels show two-dimensional maps of the ACF significance, while the right panels display the corresponding three-dimensional representations. At the native angular resolution, the ACF, which closely traces the dirty beam, exhibits highly significant sidelobes surrounding the central peak, including prominent negative sidelobes. After smoothing to $\theta_{1/2} = 1\farcs0$, these sidelobes are substantially suppressed, and the ACF becomes more compact and centrally concentrated. The suppression of the dirty-beam sidelobes reduces the spatial correlations in the image noise, thereby decreasing the number of statistically significant noise fluctuations and lowering the fraction of spurious source detections in the lower-resolution image.

Tsukui et al. 2023)[2]. Figure 5 presents both the two-dimensional and three-dimensional representations of the ACF for the two angular resolutions.

The high-resolution image exhibits prominent negative sidelobes surrounding the central ACF peak, demonstrating that the image noise is strongly correlated over spatial scales comparable to the synthesized beam. These negative bowls and sidelobes are imprinted throughout the image, producing localized negative fluctuations that are readily identified as spurious sources when the image is inverted. In contrast, after smoothing the image—either through image-plane convolution or *uv* tapering—the ACF becomes significantly smoother, with the negative sidelobes largely disappearing. The same behavior is evident in the two-dimensional ACF maps, where the high-resolution image displays pronounced, symmetric sidelobe patterns, whereas the lower-resolution image is characterized by a single, well-behaved central peak.

These results explain why the number of spurious detections is systematically underestimated by the independent-beam approximation. This approach assumes that the number of statistically independent samples is determined solely by the area of the synthesized "clean" beam, neglecting the spatial correlations introduced by the sidelobes of the dirty beam (Tsukui et al., 2023). In contrast, empirical estimates derived from inverted images naturally account for these correlations, providing a more realistic assessment of catalog reliability in sensitivity-limited surveys with approximately symmetric noise distributions.

**Highlight Concept Box**

High angular resolution does not inherently produce more spurious sources. Rather, the excess of false detections arises because the sparse *uv* sampling typically required to achieve very high angular resolution generates a point-spread function with strong sidelobes, introducing correlated noise that increases the incidence of false detections. The larger number of independent resolution elements and residual primary-beam imaging effects further contribute to the observed spurious-source fraction but cannot account for the full excess.

Future investigations should explore whether alternative imaging and deconvolution algorithms (e.g., Wiaux et al., 2009; Schmidt et al., 2022) can mitigate the impact of sparse *uv* sampling on the generation of spurious sources, which will be particularly important for future deep continuum surveys with facilities such as the next-generation Very Large Array and the Square Kilometre Array. While their enhanced angular resolution will enable unprecedented studies of galaxy structure and evolution, achieving reliable source catalogs will also require sufficiently dense *uv* coverage to suppress PSF sidelobes and minimize correlated noise. Otherwise, the improvements in angular resolution may be tarnished by an increased fraction of spurious detections in blind source catalogs.

[2] https://github.com/takafumi291/ESSENCE

## 5. FACILITIES

Karl G. Jansky Very Large Array

## 6. ACKNOWLEDGEMENTS

E.F.-J.A. and M.V.G. acknowledge support from UNAM-PAPIIT project IA102023, IA104725 and from CONAHCyT Ciencia de Frontera project ID: CF-2023-I-506. The National Radio Astronomy Observatory is a facility of the National Science Foundation operated under cooperative agreement by Associated Universities, Inc. We thank Dillon Dong, Leindert Boogaard, and Takafumi Tsukui for their helpful comments and insightful discussions.